\documentstyle[12pt]{article}

\newcommand{\F}{\noindent}

\newcommand{\MP}{\medskip}
\newcommand{\BP}{\bigskip}

\newcommand{\HH}{{\cal H}}
\newcommand{\UU}{{\cal U}}

\newcommand{\beq}{\begin{eqnarray}}
\newcommand{\ene}{\end{eqnarray}}

\Large

\begin{document}

\begin{center}
\Large

{\bf What is and What should be Time?}
\normalsize

\vskip12pt

Hitoshi Kitada

Department of Mathematical Sciences, University of
Tokyo

Komaba, Meguro, Tokyo 153, Japan

E-mail: kitada@po.iijnet.or.jp
\vskip6pt

and

\vskip6pt

Lancelot R. Fletcher

E-mail: lance@freelance.com
\vskip12pt

March 16, 1996

\end{center}

\vskip24pt

\normalsize

\leftskip24pt
\rightskip24pt

\small

\noindent
{\it Abstract}: 
The notions of time in the theories of Newton and Einstein 
are reviewed so that certain of their assumptions are clarified.
These assumptions will be seen as the causes of the incompatibility 
between the two different ways of understanding time, and seen to be
philosophical hypotheses, rather than purely scientific ones.
The conflict between quantum mechanics and (general) relativity 
is shown to be a consequence of retaining the Newtonian conception
of time in the context of quantum mechanics. As a remedy for 
this conflict, an alternative definition of time -- earlier 
presented in Kitada 1994a and 1994b -- is
reviewed with less mathematics and more emphasis on 
its philosophical aspects. Based on this revised
understanding of time it is shown that quantum mechanics and
general relativity are reconciled while preserving the
current mathematical formulations of both theories.

\leftskip0pt
\rightskip0pt

\vskip 24pt

\large
\noindent
{\bf Introduction}

\vskip12pt

\normalsize

\noindent
Previous papers of Kitada 1994a, 1994b proposed an approach to
the problem of overcoming the apparent inconsistency of
non-relativistic quantum mechanics and general relativity. The
purpose of this paper is to explain the structure and background
of that approach, with emphasis on a certain philosophical problem
related with the notion of time.

The inconsistency of quantum mechanics
and general relativity, when 
looked at mathematically, seems at first sight obvious and
inescapable from the fact that the geometry of quantum mechanics
is Euclidean, while general relativity employs a curved,
Riemannian geometry.
 
The proposal of Kitada 1994a to overcome this
apparent mathematical incommensurability of these two geometries is
by ``orthogonalizing" them; {\it i.e.} by expressing them as a direct
product $X\times R^6$, where $X$ represents the curved
Riemannian manifold associated with general relativity, and
$R^6$ (or in the usual space-time context, $R^4$) denotes the
Euclidean space of phase space coordinates $(x,v)$ of
non-relativistic quantum mechanics. As two components of the
orthogonalized total space $X\times R^6$, the Riemannian space
$X$ and the Euclidean space $R^6$ are compatible with no
contradiction.

General relativity and quantum mechanics are the two most
important and comprehensive theories of contemporary physics. By
``comprehensive" we mean that both theories claim to apply to
everything. In practice it may seem as if these two theories
applied to different physical domains, since the most striking
applications of quantum mechanics occur when we consider things
that are extremely tiny in relation to ourselves -- things like
electrons and photons -- while the most striking applications of
general relativity occur in connection with extremely large and
dense concentrations of matter and enormous spatio-temporal
magnitudes. But, in principle, every physical thing must be
capable of being described adequately by both theories, at
least this is what the theories claim. And there are certain
cases -- of particular interest in recent cosmology and
astrophysics -- where the extremes of density that are the
particular province of general relativity coincide with the
extremes of minuteness that are the special province of quantum
mechanics. In those situations, the physicist is compelled to
face a problem which is present in the background of science all
the time but which can otherwise be evaded without practical
consequence: the fact, namely, that these two comprehensive
theoretical structures appear to be mutually incompatible, that
they seem to involve different -- and contradictory --
assumptions about the nature of space, time and causality.

Our intention in this paper is to outline an approach to the
understanding of general relativity and quantum mechanics in
which these theories will appear as distinct but systematically
coordinated perspectives on the same reality. The
orthogonalization of the spatial foundations of the two theories
allows us to speak of the two theories as distinct. To express
the possibility of their systematic coordination will require a
more extended analysis of the nature of time.

In brief, we can express our approach as follows:

\begin{enumerate}
	
\item We begin by distinguishing the notion of a local system
consisting of a finite number of particles. 
Here we mean by ``local" that
the positions of all particles in a local system are understood
as defined with respect to the same reference frame.

\item In so far as the particles comprised in this local system are
understood locally, we note that these particles are describable
only in terms of quantum mechanics. In other words, to the
extent that we consider the particles solely within the local
reference frame, these particles have only
quantum mechanical properties, and cannot be described as
classical particles in accordance with general relativity.

\item Next we consider the center of mass of a local system.
Although the local system is considered as composed of particles
which -- as local -- have only quantum mechanical properties, in
our orthogonal approach we posit that each point $(t,x)$ in the
Riemannian manifold $X$ is correlated to the center of mass of
some local system. Therefore, in our approach, the classical
particles whose behavior is described by the general theory of
relativity are {\it not} understood as identical with the ``quantum
mechanical" particles inhabiting the local system -- rather the
classical particles are understood as precisely correlated only
with the centers of mass of the local systems.

\item It is important to recognize that the distinction we are
making between local systems and classical particles which are
the centers of mass of local systems is not a simple distinction
of inclusion/exclusion. For example, we may consider a local
system containing some set of particles, and within that set of
particles we may identify a number of subordinate ``sublocal"
systems. It would seem that the centers of mass of these
sublocal systems must be ``inside" the local system as originally
defined, but the sublocal system is at the same time a local
system, and we have said that the centers of mass of local
systems	are correlated with classical particles whose behavior
is to be described in terms of relativity theory.

\end{enumerate}

\F
The paradox is avoided by noting that the distinction we are
making is a distinction of reference frame, not a distinction of
inclusion or exclusion. When we speak of classical particles (or
centers of mass) we are speaking of the particle in terms of the
observer's time, which is understood as distinct from that of
the particle observed. To the extent that the time of the system
$L$ itself is adopted as the reference time, then we are
speaking of the behavior of a local system whose development
must be described in terms of quantum mechanics.

It is our contention that time necessarily has two quite
different aspects, in relativity theory, on the one hand, and in
quantum theory on the other, and the intention of this paper is
to show that these two aspects of time are in fact complementary
and that the notion of local time, which we have associated with
the quantum mechanical local system, is not only the main
ingredient of a unification of quantum and relativity theories,
but that this actually is necessary to constituting the time of
relativity theory.

The ``orthogonalization" of the geometries of quantum mechanics
and general relativity 
gives two ways of expressing the same reality so that
it means that two aspects of quantum and relativity theories
are complementary, but it
does not immediately specify the
relationship between quantum mechanics and general relativity as
branches of physics. This relationship will come to light as we
investigate the nature and conditions of observation.

We take the following stand on the property of observation,
which gives a distinction between classical and quantum mechanical
observations as well as a relation between two aspects of
nature:

\MP

\leftskip24pt
\rightskip24pt

\noindent 
The quantum mechanical, non-relativistic effects are usually hiding
 themselves from the observer, and the observation usually reveals
 classical aspect of nature, if the observer observes the centers of
 mass of {\it infinite} number of local systems outside the observer
 in accordance with {\it the observer's} own time $t_O$.
 The quantum mechanical nature of the local system $L$ appears,
 when the time $t_L$  of a fixed observed system $L$,
consisting of {\it finite} number of particles, is adopted as
 a reference time. Then the observed values of physical quantities
 of $L$ are obtained by correcting the {\it non}-relativistic
 quantum mechanical values of $L$ in accordance with
 the {\it relativistic transformation} of coordinates
 from the coordinates of the system $L$ to the observer's coodinates.

\leftskip0pt
\rightskip0pt

\MP

To state this in another way, our basic assumption is that there are
 two kinds of observation. Let us decompose the total universe,
 {\it i.e.} the set of all particles $\{1, 2, 3, ...\}$
 into a disjoint sum of the subsets $L_j$  of $\{1, 2, 3, ...\}$.
 Some of the $L_j$  may be an infinite set, but in that case
 we assume that such infinite $L_j$  does not constitute
 an observable local system. What we call local system is
 the system consisting of finite number of particles.
 Let $L_1,\cdots,L_\ell$  ($\ell$ may be {\it infinite})
be the totality of the observable
 local systems in the decomposition $\{L_j\}$
  of $\{1, 2, 3, ...\}$. Suppose that $L=(L_1,\cdots,L_k)$
 $(k(\le\ell)$ being {\it finite}$)$
constitutes a local system consisting of the particles which belong 
to some of $L_1,\cdots,L_k$. Then there are two kinds of observation.

        The first kind of observation
  is the observation of the centers of mass of the
 local systems $L_1,\cdots,L_\ell$, when the sum of the sets 
$L_1, \cdots,L_\ell$  is equal to $\{1, 2, 3, \cdots\}$,
 hence $\ell$ is {\it infinite}.
 In this case, we assume
 by axioms 4 and 5 in section V below that the observer observes
 classical phenomena about the centers of mass of  $L_1,\cdots,L_\ell$.

        The second kind of observation
  is the observation of the {\it inside} of a local system
  $L=(L_1,\cdots,L_k)$ ($k$ is {\it finite}).
 In this case, $L$  follows quantum mechanics with respect to
 the local time $t_L$  of  $L$.
When the observer {\it inquires} into the inside of
 the quantum mechanical local system  $L$,
 we assume by axiom 6 in section VI that the observer
 makes an ``ideal" decomposition of $L$  into its sublocal
 systems  $L_1,\cdots,L_k$, say, and makes an ``ideal"
 observation on the ideal classical particles $L_1,\cdots,L_k$,
 {\it i.e.} the centers of mass of  $L_1,\cdots,L_k$.
 We assume by axiom 6 that in this case, the 
quantum mechanical quantities of $L_1,\cdots,L_k$
  are transformed to classical relativistic ones
in observer's coordinates,
 obeying the general relativistic transformations
 of coordinates. During this ``ideal" observation,
 the Euclidean structure of $L$  disappears for the time being,
 and we consider the observation as an observation performed
in an ideal Riemannian manifold
 associated to the centers of mass of  $L_1,\cdots,L_k$
within the observer's coordinate system.
The quantum mechanical quantities of $L_1,\cdots,L_k$
   transformed to classical relativistic ones, 
{\it e.g.}, times, energies, and so on, of the local systems
$L_1,\cdots,L_k$, then are used to reconstruct the Hamiltonian 
$H_L$  for the local system $L=(L_1,\cdots,L_k)$
  so that the relativistic effect
 like gravitation is included in the resultant Hamiltonian  $H_L$.
 This modified Hamiltonian $H_L$  explains the relativistic
 quantum mechanical phenomena about the local system
  $L=(L_1,\cdots,L_k)$,
 when they are observed in the observer's coordinate system.

        The first kind of observation explains the classical nature
 of the phenomena which are usually observed in daily activities,
 and the second kind of observation explains
 the relativistic quantum mechanical phenomena
like the behavior of electrons accelerated near the speed of light.
Several examples of these kinds of observations will be given in
 sections VI, VII, and VIII.

We call the time $t_L$ of a local system $L$ the
 {\it local time} of the system $L$.
This notion of local time is a main ingredient of our consistent
unification of quantum and relativity theories.

In this respect,
the notion of time will be reflected first below, going back to its
definition. We will recall the notions of time in Newton's context
and in Einstein's context, and clarify philosophical difference
between them. Then we will give an explanation 
of our definition of time, which is based on some philosophical 
assumptions stated in sections III and IV, assumptions 
intermediate between
Newton's and Einstein's. We will see 
in sections V and VI how our notion of time works for the purpose
of our unification of quantum mechanics and general 
relativity.

\vskip22pt

\large
\noindent
{\bf I. Assumptions of Newton and Einstein on the Notion of Time}

\vskip12pt

\normalsize

\noindent 
Isaac Newton specifies the notion of time as follows in his {\it Principia},
 Newton 1962, p.6:

\begin{quotation}

\F
Absolute, true, and mathematical time, of itself, and from its own nature,
 flows equably without relation to anything external, and by another name
 is called duration: relative, apparent, and common time, is some sensible
 and external (whether accurate or unequable) measure of duration by
 the means of motion, which is commonly used instead of true time; such
 as an hour, a day, a month, a year.

\end{quotation}

\F
Also in pp.7-8, he states:

\begin{quotation}

Absolute time, in astronomy, is distinguished
 from relative, by the 
equation or correction of the apparent time. 
For the natural days are truly
unequal, though they are commonly considered as equal,
 and used for a
measure of time; astronomers correct 
this inequality that they may measure
the celestial motions by a more accurate time.
 It may be, that there is
 no such thing as an equable motion,
 whereby time may be accurately measured.
 All motions may be accelerated and retarded,
 but the flowing of absolute
 time is not liable to any change. The duration or
 perseverance of the
 existence of things remains the same,
 whether the motions are swift or slow,
or none at all: and therefore this duration ought
 to be distinguished from
 what are only sensible measures thereof;
 and from which we deduce it,
 by means of the astronomical equation. 
The necessity of this equation, for determining
 the times of a phenomenon,
 is evinced as well from the experiments of the pendulum clock, as
 by eclipses of the satellites of Jupiter.

\end{quotation}

\noindent
The main point of this famous
passage is to assert the existence of an absolute, true time.
However, it is important to note that Newton asserts the
existence of his absolute time by means of a distinction. There
is absolute time, which flows without reference to anything
external, and then there is relative, apparent, or common time,
which is a measure of duration made by comparison of motions.
Not only that, but although there may be no absolutely regular
motion by means of which absolute time may be accurately
represented, absolute time is an ideal standard by means of
which relative or common time is ``corrected."

Einstein's theory of relativity, as is well-known, sharply
contrasts with Newton precisely on the question of time and
space: Einstein's theory makes no reference to either absolute
time or absolute space. Einstein retains the relative or common
time which can be measured and determined by means of actual
clocks associated with each local observer, but he completely
jettisons Newton's notion of an absolute time flowing equably
for all observers.

For example, in Chapter IX of Einstein's 1920 popular
presentation of special relativity, Einstein has been exploring
whether two events which are simultaneous with respect to an
embankment next to a railway track are also simultaneous for an
observer riding in the train that is moving on the track, and
his analysis comes to the following conclusion:

\begin{quotation}

\noindent
Events which are simultaneous with reference to the embankment
are not simultaneous with respect to the train, and vice versa
(relativity of simultaneity). Every reference-body (co-ordinate
system) has its own particular time; unless we are told the
reference-body to which the statement of time refers, there is
no meaning in a statement of the time of an event.

 Now before the advent of the theory of relativity it had
always tacitly been assumed in physics that the statement of
time had an absolute significance, {\it i.e.} that it is
 independent of the state of motion of the body of reference.
 But we have just seen that this assumption is incompatible with
 the most natural definition of simultaneity; if we discard this
 assumption, then the conflict between the law of the
 propagation of light in vacuo and the principle of relativity
 (developed in Section VII) disappears.
 
\end{quotation}

\F
In this passage it is clear that, with the advent of the special
theory of relativity, Einstein has abandoned Newton's notion of
 an absolute or true time.

According the Einstein, each observer observes within his own
frame of reference -- and with respect to time his time frame is
 his own clock. The theory of relativity gives us procedures for
 coordinating observations made in diverse reference frames,
 frames that are in relative motion or even, in the case of
 general relativity, mutually accelerated.

In the context of Einstein's theories, the universe as a whole
 has no frame of reference of its own. 
Time of the entire universe
 is merely time observable by the observer.
We may imagine a way
to construct a frame of reference that is large enough to
include all observers, at least all the observers we can think
of. But no matter how inclusive this frame of reference, it is
still an essentially local framework, or an extrapolation and
coordination of local frameworks. Time for Einstein never ceases
 to be connected to the clocks of observers. It is never the
 ``true time." And time for Einstein can never be said to ``flow
 equably" for all observers, no matter how situated with respect
 to one another. The observed universe may be said to exist, but
 its observation is always connected to the reference frames of
 its observers.

	In either standpoint of Newton or Einstein, it is noticeable
 that they both made some assumptions on the {\it absolute}
 space and time.
 Newton assumed that they exist and assumed implicitly that they
 should coincide with the actual, common space and time when one deals
 with the motion of bodies. On the contrary, Einstein assumed that
 such absolute space and time do not exist, or at least need not be
 considered because they cannot be perceived actually by observation 
activities. Instead Einstein took the standpoint that what exist are
 the actual clocks and rules by which the observer can measure the
 motion of the observed systems or bodies.

	It should be noticed that this sort of assumptions can be
 made in any other different ways from Newton's or Einstein's,
 whenever the assumption can be introduced,
 resulting no inconsistency
 to the physics theory. We will present one set of such assumptions
in sections III and IV, and adopt them 
as our basic philosophical assumptions
which replace the assumptions of Newton's and Einstein's.

\vskip24pt

\large
\noindent
{\bf II. Conflict between Quantum Mechanics and Relativity}
\normalsize

\vskip12pt

\noindent
The direct motivation of the introduction of
 the special theory of relativity
by Einstein was the contradiction of Newtonian mechanics
 with  electromagnetic
theory, which was developed in 19th century. Einstein's research of 
simultaneity quoted above solved the inconsistency
 with direct alteration 
of the basic notion of simultaneity and time.
 The theory was further 
extended by Einstein to the general theory of relativity
 to include the
gravitational phenomena.

On the other hand, M. Planck introduced the notion of quantum in
 1900 in relation with the explanation of blackbody spectrum. This led 
to the introduction of the quantum mechanics by Heisenberg 1925 and 
Schr\"odinger 1926. In quantum mechanics, time plays a 
special, absolute role as seen in Schr\"odinger equation:
$$
\frac{\hbar}{i} \frac{\partial}{\partial t}\psi(x,t)+H\psi(x,t)=0,\quad
\psi(x,0)=\psi_0(x),
$$
where the Schr\"odinger operator or the Hamiltonian $H$ of the system
 is defined by
$$
H\psi(x,t)=-\frac{\hbar^2}{2m}
\sum_{j=1}^3\frac{\partial^2\psi}{\partial x_j^2}(x,t)
+V(x)\psi(x,t).
$$
Thus the solution of the Schr\"odinger equation is given by
$$
\psi(x,t)=\exp[-itH/\hbar]\psi_0.
$$ 
In this context, the time $t$ is given {\it a priori}, and then the motion
$\psi(x,t)$ of the system is derived from the Schr\"odinger
 equation by using the time evolution  $\exp[-itH/\hbar]$ of the system.

 The speciality of the role of time can be seen also by looking
 at the alternative formulation of quantum mechanics by Feynman 1948.
See also Kitada 1980 for the relation between the classical mechanics
and quantum mechanics researched along the line given by Feynman 1948.

	Because the evolution in the framework of
non-relativistic quantum mechanics is governed by
the Schr\"odinger equation, the space and time are intrinsically Newtonian
 in quantum mechanics
 in the sense that the form of Schr\"odinger equation
is not invariant with respect to the relativistic transformation
of coordinates.
 Thus quantum mechanics is not consistent with either special
 or general relativity, even if one ignores the more fundamental
 problem of relating the quantized magnitudes of the one with
 the continuous magnitudes of the other. Further, the
 discrepancies found in experiments, such as the motion of
 electrons accelerated to a velocity close to the speed of
 light, indicate that the quantum mechanics may be imperfect.
	 
 Many attempts have been made to reconcile quantum mechanics
 with relativity theories, {\it e.g.} the Dirac equation, quantum
 field theories, quantum gravity, and so on. Except for the case
 of the Dirac equation, where the special relativistic
 space-time is successfully formulated in harmony with quantum
 mechanics while retaining the special role of the time
 parameter, all of these attempts suffer from inherent
 difficulties. See, {\it e.g.}, the final sentences in the last
 section 81 of Dirac 1958. See also Sachs 1986\footnote{This
 book contains an attempt at unifying the relativity and quantum
 theories from a standpoint that is quite different from the one
 that we have taken. He starts by taking his stand within the
 classical context, and develops a matter field theory of inertia,
 which tries to realize the Einstein's ideal of a unified field
 theory.}, Chap. 2, and 1988, Chap. 10, where a comprehensive
 list of discrepancies between quantum and relativity theories
 is presented.

 These attempts have been continued until quite recently and
 have in general tended to disregard the fundamental
 discrepancy, mentioned above, between the Newtonian conception
 of time presupposed by Quantum Mechanics and the quite
 different notion of time presupposed by Einstein's relativity
 theories. Recently some useful reflections have been made on
 this point (see Isham 1993 for some review, see also Unruh
 1993, Hartle 1993). In general, however, it appears that most
 other researchers are approaching the problem of time as a
 problem of a technical nature whose solution is to be found by
 means of technical adjustments within the frameworks of quantum
 field theory, quantum gravity, etc.

\vskip24pt

\large
\noindent
{\bf III. What should be the Notion of Time?}
\normalsize

\vskip12pt

\noindent
We have stated that
the cause of the conflict 
between quantum mechanics and relativity
is in the choice of the notion of time: In quantum mechanics, time is an
 absolute notion that retains the Newtonian conception of time; while, in
 relativistic theories, time is a notion whose values are determined only
relatively.

	What then should be time, if one is given this conflict and cannot
 be satisfied with this situation of physics? One would agree that a
 certain notion of time must, at least, be able to reconcile quantum
 mechanics and general relativity, for both theories have overwhelming 
evidences and show the power of explanation of experimental and
 observational facts with outstanding accuracy.

	There are several attempts presently such as quantum gravity 
(see, {\it e.g.}, Ashtekar-Stachel 1991, Isham 1993 for these attempts)
 toward a reconciliation of quantum mechanics and general relativity,
 where the problem of time is recognized central. In many of these attempts,
 the problem is considered to be a technical one to find out how to
 identify time in relation with the quantization. There are even other
 approaches, where time does not play any fundamental roles 
at all. Several difficult problems remain unsolved in these attempts.
 Almost all of these attempts toward a unification of quantum mechanics and
 relativity have been engaged in their work with the expectation that
 the relativistic invariance under the group Diff($M$) of diffeomorphisms
 of the spacetime manifold $M$, is preserved in their final `quantized'
 universe. There are several ways in getting the final quantized world; 
{\it e.g.}, one way is to identify time before quantization, another is 
to identify time after quantization, and many others (see Isham 1993).
 These attempts go back to the invention of quantum field theory by Dirac
 in 1927. These attempts, however, have not been proved to be consistent 
(see, {\it e.g.}, Fr\"ohlich 1982 for the inconsistency proof 
of one of such 
theories), even though they can give partial explanations of some
 relativistic quantum mechanical phenomena including Lamb shift. 

	These attempts seem to be overlooking the possibility that the problem
 may be at the more fundamental level, namely at the level of philosophical
 confrontation between Newton and Einstein in their attitudes toward the
 ``absoluteness" of the total, entire universe and time. As we have stated,
Newton holds the notion of absolute time so that it could be identified
 with the accurate measure of motions to control the total universe in
 a {\it visible} or {\it observable} manner. 
Einstein claims that what actually observable are
 the relative time and coordinates, and there is no necessity for such
 an absolute time to be dominating throughout the whole universe. 

	Seeing this, we notice that this conflict disappears, if we cease
 to adhere to the `visible' absolute time of Newton, namely if we assume
 that the total absolute time {\it exists}
 but is {\it unobservable} to the observers,
 and that the {\it observable} facts are only the relative
 {\it local} motions, places, and times. Specifically, there is a room
 for reconciling this conflict, where we distinguish between the absolute
 time and the relative time so that they are, respectively, 
unobservable and observable, and have different values. Then we can hold
 the notion of absolute time for quantum mechanics,
 without contradicting the actually observable (general) relativistic time.

	To state this assumption in other words:

\begin{quotation}

\F
{\it What the observer can see is the local classical 
(general) relativistic 
phenomenon. On the contrary, 
the total universe is quantum mechanical, and 
its quantum mechanical nature cannot
 be seen directly by any observer.}

\end{quotation}

\F
This assumption or standpoint is one possibility by which one can avoid
 the above-mentioned philosophical discrepancy between Newton and Einstein.
 Even if one adopts this standpoint, there might be several passages to
 proceed. We here present one of the possible passages.

\vskip20pt

\large
\noindent
{\bf IV. An Alternative Notion of Time}
\normalsize

\vskip12pt

\noindent
In this and subsequent sections, we state an outline of our basic framework
 of one possible way to reconcile the quantum mechanics and general
 relativity. In doing so, we avoid the rigorous mathematical arguments
 as far as it is possible, except in some places where we need to clarify 
the precise meaning for those who might feel it necessary. We also avoid
 the logically firm formalism in order to make the meaning of our theory
 clear. It would be, however, helpful to the reader to remark that our
stand of the following description is basically the mathematical
formalism: Our framework consists of six axioms, axiom 1 through 6
stated in the followings in some implicit way, which are assumed as
our basic postulates that must hold {\it a priori} in our physical world
and none of which can be derived from the other axioms.

\BP

	Our basic assumption as an alternative for Newton's absolute time is
 that the total, entire universe has {\it no}
 time. Namely, contrary to Newton's absolute time that flows equably
 throughout the entire universe, we assume that the total universe is
 static and stationary. More specifically, the total universe is assumed 
to be a quantum mechanical bound state, {\it i.e.} an eigenstate of a
 Hamiltonian, denoted $H$, of infinite degrees of freedom, in a certain sense
 (see axiom 1 in Kitada 1994a, 1994b). (We should state that our notion
 of the eigenstate of $H$ with infinite degrees of freedom is a rather
 weak one than the usual meaning of eigenstate.)\footnote{If we try to be
rigorous in philosophical sense, we should
remark that the non-existence of time in our context should be
interpreted as the ``eternity" in Spinoza's sense (Spinoza, {\it The Ethics}
in Descartes, Spinoza, and Leibniz 1960)
rather than as the conventional meaning of the eternal continuance 
that lasts forever without a beginning or end. In Definition VIII 
of {\it The Ethics}, Part I, Spinoza states
\begin{quote}

VIII. By {\it eternity}, I mean existence itself, in so far as it is 
conceived necessarily to follow solely from the definition of that which
 is eternal.

{\it Explanation.} --- Existence of this kind is conceived as an eternal 
truth, like the essence of a thing, and, therefore, cannot be explained
 by means of continuance or time, though continuance may be conceived 
without a beginning or end.

\end{quote}

\noindent
Our axiom 1 which asserts that the total universe, which will
 be denoted $\phi$, is an eigenstate of a total Hamiltonian $H$,
 means that the universe $\phi$ is an eternal truth,
 which cannot be explained in terms of continuance or time.
 In fact, being an eigenstate contains
 no notion of time as seen from its definition: $H\phi=\lambda\phi$ for
 some real number $\lambda$. 
The reader might think that this definition just states
that the entire universe $\phi$ is freezing at an instant which lasts
 forever without a beginning or end.
 However, as we will see, the total universe $\phi$
 has an infinite degrees of freedom inside itself, as internal motion of
 finite and local systems, and never freezes. Therefore,
as an existence itself, the universe $\phi$ does not change, however, 
at the same time,
it is not freezing internally.
These two seemingly contradicting aspects of the universe
$\phi$ are possible by the quantum mechanical nature of the definition
of eigenstates.}

	Thus the universe itself does not change. However,
inside itself, the universe can vary quantum mechanically,
 in any local region or in any local system consisting of a finite 
number of (quantum mechanical) particles. Therefore, we can define a
 {\it local time} in each local system as a measure or a clock of
 (quantum mechanical) motions in that local system. 

	In other words, for a local system $L$ with $N$ number of particles 
$1, 2, 3, ..., N$, there can 
be defined the position vectors $x_1,x_2,x_3,\cdots,x_N$
  and momentum vectors $p_1=m_1v_1,p_2=m_2v_2,p_3=m_3v_3,\cdots,
p_N=m_Nv_N$, where $m_j$  is the mass of the $j$-th particle, so that 
the correspondent quantum mechanical selfadjoint operators $X_j=(X_{j1},
X_{j2},X_{j3})$  and $P_j=(P_{j1},P_{j2},P_{j3})$
  in a Hilbert space $\HH=L^2(R^{3n})$ of $N=(n+1)$ particles, satisfy
 the so-called canonical commutation relation. (This statement is axiom 2
 of Kitada 1994a.) Then the local time $t_L$ associated with the local system
$L$ is defined as a quotient of position $x_j$  by velocity $v_j=p_j/m_j$
\beq
t_L=\frac{|x_j|}{|v_j|}.
\ene
Here we note that the right hand side of this definition looks depending
 on the number $j$. But it is known (Enss 1986) that it does not depend on
 $j$, if one defines the right hand side in a certain quantum mechanical
 way as in Kitada 1994a, sections 4-5 (see axiom 3, Theorem 1, and 
Definitions 1-3 there, and the paragraph after the formula (3) 
below for more precise descriptions). Thus local time is defined
 as a {\it measure of motion} inside each local system.
This notion of local times is not contradictory with the nonexistence of
the total time: The total universe of infinite number of particles
is stationary as will be described in section V below. Any local system
 of finite number of particles, however,
 can be nonstationary, and can vary inside
itself, as a consequence of the variation outside the local system, which
compensates the change inside the local system, so that the stationary
nature of the total universe is preserved.

	We remark about the difference between our definition of 
local times and the conventional understanding of the notion of time. 
The common feature of the conventional understanding of time, 
including Newton's definition of absolute time, is that the time 
is something existing or a one given {\it a priori}, independently of
 any of our activities, {\it e.g.} of observation activities. 
In our definition, time is not an {\it a priori} existence, but 
a convenient measure of motions inside each local system. 
Our definition of local times mentioned above at the end of
the paragraph before formula (1) is that a local time is a clock
--- which measures, not time, but the motions of the local system.
Differently from the conventional understanding where time is
given {\it a priori}, the clock does not {\it measure} time,
but it {\it is} time in our definition. Further, as we will
state in section V in the paragraph after formula (3), the proper clock
is the local system itself, and it is a necessary manifestation
of that local system. In this sense, ``clocking" is
the natural activity of any local system. It follows from
this that to be an existing thing in the world necessarily
involves clocking, without which there is no interaction.
In these senses, our stand is in complete contradiction to
the conventional understanding of time measurement, where
time is given {\it a priori} and clocks measure those times, therefore
the measurement of time is an incidental activity. Contrary to the
conventional understanding, our stand is that all beings are
engaged in measuring and observing, and the activities of measuring and
observing are not incidental, but pertain to the essence of
all interactions. If we are permitted to express it somewhat boldly,
we have turned things completely around: It is not that things exist
and their duration is incidentally expressed by clocks. According to our
formulation, clocks exist and their operation is necessarily
expressed by duration.

We explain these in a more physical manner. As
 noted above, the formula (1) fortunately defines a common parameter
 $t_L$ associated to each local system $L$, because it is independent
 of the particular choice of the particle number $j$. This is exactly
 a magical fact  which enables one to define time, as well as, that
 makes one believe that time is an existence which exists outside
 us {\it a priori}. However, the formula (1) holds only in an 
approximate sense as explained in p. 288 of Kitada 1994a, related
 with the uncertainty principle, and there is a possibility that
 the difference among the quotients $|x_j|/|v_j|$ $(j=1,2,\cdots,N)$ 
can be detected by experiments (see Kitada 1994b). What looks 
{\it a priori} about the existence of the time coordinate is only 
a disguise in this sense. Instead, what exist {\it a priori} are 
positions and motions (velocities), through which the approximate
 values of time can be measured by taking the quotients $|x_j|/|v_j|$ 
in each local system $L$ consisting 
of $N$ particles $j=1,2,\cdots,N$. Time is,
 therefore, a quantity determined approximately by experiments by
 the use of formula (1), and is not an existence that is {\it a priori}.
We will touch on this point again in section V.

	Nevertheless, once a local time is identified by the formula (1),
 as a measure of motion, as
$t=|x|/|v|$ in each local system, our definition of local times is a
 specification or a clarification of the `relative, apparent, and common
 time' measured `by the means of motion, which is used'
 `instead of true time' in Newton's sense (see the first quotation from 
{\it Principia}, Newton 1962). It is a realization of Einstein's local 
nature of time and coordinates, as well, in the sense that the local time is
defined only for each local system consisting of a finite number of
 particles.

	We next remark on the relation of our notion of time with those of
Newton's and Einstein's.
Newton's  assumption in {\it Principia}
is the existence of `true time' flowing equally throughout
 the universe to give an absolute and accurate meaning to the `usual,
 common time,' by identifying his absolute time with an idealization of
 the common time. This assumption was necessary for 
the unified treatment of the motion
 of bodies which was considered common throughout the universe.

	But we know from the proposal of Einstein's special relativity that the
 absolute time flowing equably throughout the entire universe is
 unnecessary in constructing physics theory. We know, even more, from
 Einstein's proposal that such a notion would do more harm than 
good in constructing a {\it consistent} physics theory.

	In this point our choice is to follow Einstein in abandoning the notion
 of equably flowing time, controlling the entire universe,
 so that we adopt the local times
 for describing local motions, that will be shown to be consistent with 
the general relativity.

	On the other hand, we also follow Newton in the point that it is
 necessary to consider the entire `true' time in order to have a synthetic
 view to nature.
 For this purpose, we adopt, as our actual definition of the true 
time, the assumption that the entire, true time does not exist.

	To sum up, we have {\it localized} Newton's notion of time so that it
 will be shown to be {\it consistent} with the relativity. On the other hand,
 we adopted, as a substitute for Newton's total, absolute time, 
 a stationary static universe, instead of 
Newton's dynamical universe changing linearly and equably forever.

We finally remark that
the assumption that the total universe does not change,
 implies that there exists a quantum mechanical correlation
 among the local systems inside the universe\footnote{In this respect,
our stand is quite similar to the one adopted in D. Bohm and B. J. Hiley 1993.
They consider the universe as the one that cannot be divided, and is internally
correlated. Also we remark that their notion of the ``implicate order" 
(see Chap. 15 of Bohm-Hiley 1993) is essentially the same as our notion
 of ``local systems." Our local systems are determined
by their outside and the total universe itself,
to conclude the stationary nature of the universe. 
Vice versa, local systems and the stationary nature of the total universe
determine the outside of the local systems.
Bohm-Hiley's implicate order is another expression of this statement.}.
 Namely if a local system
 changes, then the effect of that change propagates instantaneously to 
other local systems inside the universe in a quantum mechanical
 way, or in other words any change in any local region of the universe is
 compensated by the corresponding change of the rest of the total universe.
 In this sense, the total universe is inseparably related inside itself 
in quantum mechanical way.

\vskip24pt

\large
\noindent
{\bf V. Independence of local systems and the Relativity}
\normalsize

\vskip12pt

\noindent
Before introducing the relativity into our context, we first remark that,
 any two local systems $L_1,L_2$  inside the total universe are not
 correlated in classical way in their local times $t_{L_1}$
  and $t_{L_2}$, although
 they are correlated inside the total universe in
 a quantum mechanical way. 
We note that this statement is not contradictory,
 because the quantum mechanical nature and classical nature
 of the universe are mutually independent aspects of the universe;
 therefore,
these local 
times correlated in quantum mechanical way
 can be noncorrelated in classical manner,
 as two mutually independent aspects of nature.
 Namely the local times $t_{L_1}$  and $t_{L_2}$, and further the
 coordinates of $L_1$  and $L_2$, are mutually independent in any classical
 manner. 

	The verification of the last statement requires some definitions in
 Kitada 1994a or Kitada 1994b. We here mention only that this independence
 of $t_{L_1}$ and $t_{L_2}$ is a consequence of our choice of the 
Hilbert space 
\beq
{\cal U}=\{\phi\}=\bigoplus_{n=0}^\infty \left(\bigoplus_{\ell=0}^\infty
{\cal H}^n  \right), \   
{\cal H}^n=\underbrace{{\cal H}\otimes\cdots\otimes{\cal H}}_{n\
\mbox{\scriptsize{factors}}}, \ 
{\cal H}=L^2(R^{3}),
\ene
of possible universes $\phi$. We briefly explain this definition.
$\phi$ represents a possible `state' of the 
total universe. The space $\HH$ of the state vectors can be regarded as
 representing a flat Euclidean space where one particle $j$ lies. For
 each particle $j$, there is a state space $\HH$, the tensor product
$\HH^n=\HH\otimes\cdots\otimes\HH$
  of whose $n$ copies represents the $N=n+1$ particle Euclidean space.
 Noting that there are infinitely many different sets of particles with
 the same number $N=n+1$ of particles, we get an infinite sum 
$\bigoplus_{\ell=0}^\infty\HH^n$  of the same space $\HH^n$.
 Then summing this space with respect to the number  $n=N-1(\ge0)$,
 we define the total Hilbert space $\UU$ of possible universes $\phi$ by
$\UU=\bigoplus_{n=0}^\infty\left(\bigoplus_{\ell=0}^\infty\HH^n\right)$.

	That the universe $\phi$  does not change, as described in section IV,
 means in this context that $H\phi=\lambda\phi$ for some real number 
$\lambda(\le0)$  in a certain sense, where $H$ is the total Hamiltonian
 defined on the space $\UU$ of possible
 universes as mentioned in section IV
(see axiom 3 in Kitada 1994a for precise definition).
 In the definition of the above
$\UU$, the subscript $n$ represents the number $N=n+1$
 of (quantum mechanical) particles of the local systems under 
consideration. (Therefore the total universe $\phi$ represents one possible
 state of the universe, consisting of {\it infinite}\footnote{To be precise
in philosophical sense,
the {\it infinity} property of the universe $\phi$ here 
should be interpreted as the ``absolutely
 infinite" in the sense of Spinoza, {\it The Ethics}, Definition VI of Part I:
\begin{quote}

VI. By {\it God}, I mean a being absolutely infinite --- that is, 
a substance consisting in infinite attributes, of which each expresses
 eternal and infinite essentiality.

{\it Explanation.} --- I say absolutely infinite, not infinite after its
 kind: for, of a thing infinite only after its kind, infinite attributes
 may be denied; but that which is absolutely infinite, contains in
 its essence whatever expresses reality, and involves no negation.

\end{quote}

\noindent
Our total universe $\phi$ is a perfection in the sense that it has no 
outside, and everything is inside $\phi$. The infinite nature of $\phi$ is,
 therefore, not an infinity after its kind in the sense of the following
Definition II of {\it The Ethics}, Part I,
 and must be the absolute infinity. 

\begin{quote}

II. A thing is called {\it finite after its kind}, when it can be limited 
by another thing of the same nature; for instance, a body is called finite
 because we always conceive another greater body. So, also, a thought is
 limited by another thought, but a body is not limited by thought, nor a
 thought by body.

\end{quote}

\noindent
In this sense,
 our universe $\phi$ is a perfection with no limits, as well, in the sense
 of Leibniz, {\it The Monadology}, 41 (Descartes, Spinoza, Leibniz 1960,
 p. 461):

\begin{quote}

41. Whence it follows that God is absolutely perfect, perfection being 
understood as the magnitude of positive reality in the strict sense,
 when the limitations or the bounds of those things which have them are
 removed. There where there are no limits, that is to say, in God,
 perfection is absolutely infinite.

\end{quote}} 
 number of particles.)
 The additional subscript $\ell$ in the double sum in  $\UU$ 
represents the difference of the particles that belong to various systems
 with the same number of particles.
Even if two local systems $L_1,L_2$ have a common part $L_3$
consisting of those particles which belong to both of $L_1$
and $L_2$, the two local Hilbert spaces $\HH_{L_1}$ and $\HH_{L_2}$
of the state functions for the particles in $L_1$ and $L_2$ are
different spaces, say $\HH^{n_1}$ and $\HH^{n_2}$, where
$n_j+1$ is the number of particles in $L_j$, hence $\HH_{L_1}$ and
$\HH_{L_2}$ are {\it mutually independent} when considered in $\UU$,
which is also the case even when $n_1=n_2$ unless $\ell_1=\ell_2$.
 This fact gives us an arbitrariness for
 the classical mechanical relation
 between the local times $t_{L_1}$  and  $t_{L_2}$
(see also p. 289 of Kitada 1994a). 
Due to this arbitrariness, {\it we have a complete freedom to make any
 assumption on the classical relation of} \ $t_{L_1}$ {\it and}
$t_{L_2}$, {\it and also on the classical relation of 
the coordinate systems of} \ $L_1$
{\it and} $L_2$, with keeping the quantum mechanical
correlation inside the total universe, as a consequence of the relation
$H\phi=\lambda\phi$. We can therefore assume an
 arbitrarily fixed transformation: 
\beq
(t^2,x^2)=f_{21}(t^1,x^1)
\ene
between the classical coordinate systems $(t^j,x^j)$ of any two local
 systems $L_j$ $(j=1,2)$. 

We summarize what we have described in a more precise manner.
A {\it local system} $L$ is a system consisting of a finite number,
say $N=n+1$, of particles. Let $\phi$ be the universe chosen from
 the set $\UU$ of possible universes. Then the state vectors 
$\psi_L=\psi_L(x_1,\cdots,x_n)$ of the local system $L$ of $N=(n+1)$ 
number of particles are obtained from the universe $\phi$ as 
$\psi_L(x_1,\cdots,x_n)=\phi(x_1,\cdots,x_n,x_{n+1},x_{n+2},\cdots)$
 with varying $x_{n+1},x_{n+2},\cdots$ as the parameters that determine
 each state $\psi_L$ of the local system $L$, corresponding to each choice
 of the infinite number of coordinates $(x_{n+1},x_{n+2},\cdots)$. 
The totality of such $\psi_L$ constitutes a {\it local universe} $\HH_L$
 which represents a state space for the local system $L$. The local
 Hamiltonian $H_L$ for $L$ is the restriction of $H$ onto $\HH_L$. 
Then the {\it proper} or {\it local clock} for $L$ is defined as the 
unitary group $\exp[-itH_L]$ $(t\in R^1)$ on $\HH_L$, and the parameter
 $t$ in the exponent of the group $\exp[-itH_L]$ is called the 
{\it proper time} or {\it local time} of the local system $L$. 
This is the precise definition of the local time of a local system, 
and gives rise to the definition (1) of the local time described above
 by the use of Enss' result, 1986. 
We notice that this procedure is a 
reverse one to the usual procedure adopted in physics to describe the 
physical phenomena, the procedure stated in section II,
 where time $t$ is given a priori,
 and only after $t$ given, the motion of the local system is described by
$\psi(x,t)=\exp[-itH/\hbar]\psi_0$, 
by utilizing the evolution  $\exp[-itH/\hbar]$, 
which in our case plays the role of the clock of the system.
In this sense, our definition of local times reverses
the conventional understanding of time, as stated in section IV.

\BP

	To make the physical meaning implied by these assumptions and
 considerations clear, we introduce here two kinds of conventional
 coordinate systems associated with each local 
system $L$. 
One is the {\it proper coordinate system} of $L$ which is the Euclidean
 coordinates describing the quantum mechanical world inside the local
 system $L$. Another is the {\it observer's coordinate system},
 which emerges when the observer {\it observes} the outside
in classical way and
 is the curved Riemannian coordinates associated with the local system 
$L$, that measures, as an observer, the classical world,
 outside the observer's local system, 
consisting of infinite number of 
classical particles. (We repeat that, by classical particles,
 we mean the {\it centers of mass} of local systems. On the other hand,
 the particles whose positions and momenta determine the local 
time (1) as described above, are quantum mechanical particles.)
 For convenience sake, we assume that, at the center of mass of a local
 system $L$, the space coordinates $x=0$  for both 
proper and observer's coordinate systems.

	In other words, the coordinate system $(t,x)\in R^4$
  of a local system $L$, defined by using 
the local time parameter $t=t_L=|x_j|/|v_j|$,
 plays double roles which are mutually independent: 
one role is the measurement of the inside of $L$ itself,
 and another is the observation of the 
{\it centers of mass}{\footnote{We here remark that the 
{\it center of mass} of a local system $L$ consisting of $N$
 number of particles $1, 2, ... ,N$ with positions 
$x_1,\cdots,x_N$ and masses $m_1,\cdots,m_N$,
is a position vector 
$$
x=(m_1x_1+\cdots+m_Nx_N)/(m_1+\cdots+m_N).
$$
The center $x$  of mass is the configuration part $x$
  of a point $(t,x)$  of the Riemannian manifold $X$.
 This is the meaning of the correlation between the center $x$
  of mass and the point $(t,x)$  in X, 
which was stated in the introduction.}}
 of other local systems $L'$, at least in a vicinity of the center
 of mass
 of $L$. (In particular, $L'$ can be equal to $L$.) These two roles
 are distinguished from each other by the difference of their 
metrics $g_{\mu\nu}(t,x)$: For Euclidean proper coordinates,  
$g_{\mu\nu}(t,x)\equiv 1$ for $\mu=\nu=0,1,2,3$  and $\equiv 0$
  for $\mu\ne \nu$, and for the curved Riemannian observer's coordinates, 
$g_{\mu\nu}(t,x)$  is a general metric tensor, under the assumption of
 the general principle of relativity adopted below. Namely $g_{\mu\nu}$
  satisfies the transformation rules:
\beq
g^1_{\mu\nu}(y_1)=g^2_{\alpha\beta}(f_{21}(y_1))
\frac{\partial f_{21}^\alpha}{\partial y_1^\mu}(y_1)
\frac{\partial f_{21}^\beta}{\partial y_1^\nu}(y_1),
\ene
where $y_1=(t^1,x^1)$; $y_2=f_{21}(y_1)$  is the transformation from
 the coordinates $y_1=(t^1,x^1)$  to $y_2=(t^2,x^2)$,
 which was fixed arbitrarily in (3) above; and $g^j_{\mu\nu}(y_j)$
  is the metric tensor expressed in the observer's coordinates 
 $y_j=(t^j,x^j)$.

	From the obvious fact that the {\it center of mass}
 of a local system $L$ is independent of the {\it inner relative}
 coordinates of $L$, these two roles of coordinate system of 
$L$, {\it i.e.} the roles as the proper and observer's coordinates, are
 independent mutually. Further, between the observers' coordinate systems
 of any two local systems $L_1$  and  $L_2$, there is complete
 arbitrariness in making any assumption on the
 classical relation, as mentioned above.

We repeat here our standpoint that the quantum mechanical world
and classical world are mutually independent as two independent
aspects of nature, expressed as a direct product $X\times R^6$.
This implies that the proper coordinate system of
 $L$ does not contradict the observer's coordinate system
 of $L$. 
To state this difference in accurate manner, the classical observation
 is concerned with the {\it infinite} number of local systems 
$L_1,\cdots,L_\ell$, which decompose the total universe $\{1, 2, 3,\cdots\}$,
and the quantum mechanical observation is with the
{\it inside} of a local system
$L$ consisting of {\it finite} number of particles. Therefore there is 
neither intersection nor overlapping between these two kinds of
 observation and between
the corresponding coordinate systems of 
proper and observer's coordinate systems.

Therefore, we can assume, as have been anticipated, the principles of
 general relativity (general principle of relativity, and the principle
 of equivalence) as a classical
 relation between the observers'
 coordinate systems of $L_1$  and  $L_2$. Explicitly, we assume the
 following two postulates:
\BP

{\it General Principle of Relativity} (axiom 4).  Those laws of physics
 which control the {\it relative} motions of the {\it centers}
 of mass of the {\it observed} local systems are expressed by the classical
equations which are covariant under the change of {\it observers'}
 coordinate systems of  $R^4$.
\BP

{\it Principle of Equivalence} (axiom 5).   The coordinate system 
$(t^1,x^1)$  associated with the local 
system $L_1$  is the local Lorentz system of coordinates. Namely,
 the gravitational potentials   $g_{\mu\nu}$
for the {\it center} of mass of the local system  $L_1$,
 {\it observed} in these coordinates  $(t^1,x^1)$, are equal to 
 $\eta_{\mu\nu}$. Here    $\eta_{\mu\nu}=0$ $(\mu\ne\nu)$,    $=1$ 
$(\mu=\nu=1,2,3)$, and $=-1$   $(\mu=\nu=0)$.
\BP

	Namely, we assume that the classical world outside an
 observer's local system obeys the general relativity with respect 
to that observer's coordinate system. 

	On the other hand, we have been assuming, from the first stage of
 our definition of local times, the quantum mechanics, for the
relative motions of the particles inside every local
 system. In other words, we assume that the world 
inside each local system obeys the quantum mechanics with respect
 to the proper coordinate system of that local system.
\MP

	We then have the following
\MP

{\bf Theorem.}  The additional postulates of general relativity for
 the outside of the observer's local system are {\it consistent}
 with the quantum mechanical postulates inside each local system. 
\MP

\F
{\it Proof:}  The general relativity is postulated on the framework of
 the observer's coordinate system, and the quantum mechanics is 
postulated within the proper coordinate system. These 
two kinds of coordinate systems are independent mutually.
 Therefore, these two postulates (or any other kinds of postulates)
 are consistent, if these postulates are consistent with the metrics 
of these coordinates. 

	In our case of general relativity and quantum mechanics,
 we first recall
 that the relation of any two of the observers' coordinate systems can
 be determined arbitrarily as stated above. Thus the observer's coordinate
 system can be chosen to be the curved Riemannian coordinates, 
{\it consistently} with the general relativity. In other words,
 the postulates of general relativity can determine the metric of the
 observer's coordinate system. Namely, recalling that $x^2=0$
  at the center of mass of a local system  $L_2$, we see that the
 principle of equivalence (see axiom 5 above) implies for
$g_{\alpha\beta}^2(y_2)$   in (4) that  $g_{\alpha\beta}^2(t^2,0)
=\eta_{\alpha\beta}$, where 
   $\eta_{\alpha\beta}=0$ $(\alpha\ne\beta)$,    
$=1$ $(\alpha=\beta=1,2,3)$, and    $=-1$ $(\alpha=\beta=0)$.
 (The consistency of this axiom 
follows from the above-mentioned independence between the center of mass of
 a local system $L$ and $L$'s inner proper coordinates.
See Kitada 1994a, Theorem 2.)
 Thus the metric at the center of mass of the local system $L_2$, 
observed from the local system  $L_1$, is determined by the above
 transformation rules (4) for   $g_{\mu\nu}^j$
as follows:
\beq
g^1_{\mu\nu}(f_{21}^{-1}(t^2,0))=\eta_{\alpha\beta}
\frac{\partial f_{21}^\alpha}{\partial y_1^\mu}(f_{21}^{-1}(t^2,0))
\frac{\partial f_{21}^\beta}{\partial y_1^\nu}(f_{21}^{-1}(t^2,0)).
\ene

	For the quantum mechanics, it is obvious that the Euclidean proper
 coordinates of a local system naturally describe the quantum mechanics
 consistently.                 Q.E.D.
\BP

	See Kitada 1994a, section 6, where another proof is given in 
Theorem 2. It utilizes the two sorts of independence mentioned above;
 the one between the observers' coordinates of any two local systems,
 and another between the center of mass of a local system $L$ and $L$'s
 inner proper space. See also Kitada 1994b, section 3, where 
a mathematical argument utilizing the notion of vector bundle is
 presented to show the above theorem.

	We make a remark that the formula (4) or (5) above gives a relation of 
the metric  $g_{\mu\nu}^j$ $(j=1,2)$ 
  between two local systems $L_1$  and  $L_2$,
 but these formulae do not determine $g_{\mu\nu}^j$ in 
any concrete sense. To determine $g_{\mu\nu}^j$  requires another
 assumption like the Einstein field equation in the usual general
 relativity. We suggest another possibility on this point: The field 
equation can be chosen arbitrarily as far as it is consistent with the
 two principles of general relativity. We may choose the Einstein 
field equation itself as one candidate. We may also adopt 
the Hoyle-Narlikar field equation (see {\it e.g.}, Narlikar 1977, Arp 1993).
 Both of Einstein's and Hoyle-Narlikar's are consistent with
 general relativity, or they are equivalent with each other in 
a certain mathematical sense (see Narlikar 1977). We will touch on this
 point again later in section VIII.

\vskip24pt

\large
\noindent
{\bf VI. Observation, that gives a connection between quantum mechanics
 and general relativity}
\normalsize
\vskip12pt

\noindent
In this way, we regard our world as a unity of the quantum mechanical
 world inside each local system and the classical world outside the
 observer's local system.  

	As stated in section II, it is commonly considered that quantum mechanics
 (QM) and general relativity (GR) are incompatible with each other.
 For instance, QM requires Euclidean space as the base space
 and the QM laws are linear, while GR the curved Riemannian space and
the GR laws nonlinear; QM is non-local and non-causal, while GR obeys
 the causality; and so on (see Sachs 1986, Chap. 2). This is the case if one
sees the two theories as the same leveled theories, or the theories on
 the same plane.

	We have, however, seen that these two theories becomes consistent
 with each other, if we orthogonalize them mutually: Set QM on a
 Euclidean space-time $R^4$  as an inner space 
tangent to the curved Riemannian space-time $X$ of classical general
 relativity. Then in the total physics space, namely in the product 
$X\times R^4$  (or  $X\times R^6$) of these spaces, QM and GR are 
mutually independent, hence mutually consistent.

	Then, are QM and GR non-interrelated entirely, if they are independent
 of each other? Quite contrarily, this independence 
is of completely logical nature and makes it possible
 to assume an arbitrary but fixed relation between these two aspects
 of nature. This relation is given by ``observation" process,
which will solve many fundamental incompatibilities between QM and GR
as listed in Sachs 1986, Chap. 2.

	In our theory, observation of the inside of
 a local system $L$ by an observer's
 local system $O$ corrects the quantum mechanical values
 associated with the particles inside the system $L$, by a 
relativistic transformation of coordinates associated with observation,
 into a relativistic quantum mechanical values. The QM and GR are
 correlated in this sense.

We here make an important remark that {\it all} possible local systems
 exist {\it a priori} by definition, {\it i.e.} all possible sets
 consisting of finite number of particles are local systems.
 When one observes other local
 systems the observer makes a choice of the local system under
 observation, but the existence of other local systems is a logical
 one and not affected. 
More precisely speaking, when an observer makes an observation, 
the observer makes or chooses one cluster decomposition of the 
total universe, {\it i.e.} it decomposes the set $\{1, 2, 3,\cdots\}$
 of all particles into a disjoint sum of subsets $L_j$ of 
$\{1, 2, 3,\cdots\}$, where $L_j$ may be an infinite set while 
in that case the particles in $L_j$  do not constitute a local system,
 hence are not observable. Then the observer observes the centers 
of mass of those local systems $L_j$.
 Thus at each moment, the observer observes those specific local
 systems characterized by this decomposition. At another moment, the
 decomposition can be changed to another one. However at each moment 
in the time frame of the observer, it observes just one set of local
 systems corresponding to a decomposition of the total universe.
 Thus although all decompositions of the universe are possible 
logically, the actually observed decomposition is just one at 
each moment for each observer. Therefore, here is no problem of 
overlapping between different observations whether or not they are 
made by the same observer. Moreover this formulation explains
 the apparent inconsistency like the EPR paradox, as the confusion
 of different observations made based on different cluster
 decompositions of the universe. We give some other examples of
 this type of paradoxes in section VII.

 Our theory can be further rephrased as an explanation
 of the daily observation activities. We do not try to explain, 
{\it e.g.}, the observation $O_2$  of the observation activity  
$O_1$, which has been one of the observation problem. Instead, 
this we explain as an observation of a physical phenomenon that 
a local system of the observer of the observation $O_1$ interacts 
with the observed system. The observation in our theory is concerned
 with only the final stage of observation in this respect.

Another important remark on the term ``observation" is that the term
is an undefined one in the sense that it is defined in axiom 4
without specific definition. To clarify this some more, 
let us take the example above of the observation $O_2$ of an 
observation activity  $O_1$.  Usually people regards $O_2$
 as an explanation of the observation $O_1$.
 Then they meet the problem of premeasurement and objectification
(see Auyang 1995, section 4). In our framework, we regard $O_1$
 merely an interaction process between the observer of 
the observation $O_1$ and the observed system, and see 
by the observation $O_2$ that $O_1$ follows Schr\"odinger equation
 with no inconveniences which the current views meet.
 As to the observation $O_2$, we thus
need not worry about what it is
 or how it should be defined or explained inside the theory.
 We just do the ``observation," which in the context of our daily 
activities has a position similar to the position that 
the undefined term ``observation" has in our theory. In this sense,
 the term ``observation" represents a final stage of observation, 
namely it is our actual deed of observation which cannot be 
analyzed any more.

\BP

Let us return to our problem of QM and GR.
	The actual relation which we impose between QM and GR
can be described as an assumption about the
 observations of the inside of an observed local system, from an
 observer's local system. We refer to axiom 6 in Kitada 1994a, 1994b
 for this assumption, which states that 

\begin{quotation}

\F
{\it the actually observed values when observing the inside of 
a local system are 
the} {\bf classical} {\it ones which are obtained by correcting the bare}
{\bf quantum mechanical} {\it values inside the observed local system,
 in accord with the relativistic change of coordinates from the observed
 local system to the observer's local system.}

\end{quotation}

\F
We remark that what is actually observable is the scattering
amplitude or differential cross section of the scattering experiment
or observation. These are the quantities observed at the final stage
of the observation activities. Other intermediate observation would
change the process itself, hence must not be done during the
experiment. 
Other quantities than the final scattering amplitudes,
 {\it e.g.}, the intermediate position and velocity of the
centers of mass are not observable directly.
Nevertheless, we assume that these observation can be done
in an ``ideal" sense, and assume the above assumption also on this
kind of observation as well as on the actual observation of
scattering amplitudes.

\MP

\F
To state more concretely this assumption:
\MP

{\it Axiom} 6.  The momenta $p_{Lj}=m_jx_{Lj}/t_{L}$  of the particles 
$j$ with mass $m_j$  in the observed local 
system $L$ with coordinate system  $(t_L,x_L)$,
 are observed, by the observer system $O$ with 
coordinate system  $(t_O,x_O)$, as  $p_{Oj}=m_jx_{Oj}/t_O$, 
where  $x_{Oj}$ is obtained from $x_{Lj}$  by the 
relativistic transformation of coordinates:
$(t_L,x_L)$ to $(t_O,x_O)$  as in axiom 4.
The same is true for the observation of the energies of the particles:
 the energies of the particles in the observed 
local system are observed by the observer as the ones transformed 
in accordance with the relativity.

\MP

	Namely, it is assumed that the quantum mechanical momenta 
$p_{Lj}=m_jx_{Lj}/t_L$  of the 
particles within the system $L$ are observed in actual 
or ideal experiments or
observations by the observer system $O$ with coordinate system  $(t_O,x_O)$,
 as the {\it classical} quantities $p_{Oj}=m_jx_{Oj}/t_O$  whose 
values are calculated or predicted by correcting the quantum mechanical
 values  $p_{Lj}$ with taking the relativistic effects of observation
 into account. A similar assumption is made for the energies 
of the particles. (See the examples below and 
in pp.296-297 of Kitada 1994a, where
it is explained how to actually `apply' these assumptions to treat the 
probabilistic nature of QM. In the latter example, the differential cross 
section $d\sigma/d\Omega$ is the {\it classically} observed value 
when the energy $E$ and the angle $\theta$ are fixed, or if we state it 
in physical terminology, are `measured' in the scattering experiment
considered there. The {\it bare} quantum mechanical value for
$d\sigma/d\Omega$ in this case is the formula (14) in Kitada 1994a, and
the {\it relativistically} corrected value is the formulae (17)
and (19).)

\MP

\noindent
{\it Proof} of the consistency of axiom 6: This axiom 6 is consistent
 with our former assumption of QM and GR, because axiom 6 is concerned
 entirely with {\it how nature looks at the observer}
when one observes the inside of a local system, and 
is independent of the logical validity of our theory.                              Q.E.D.

\MP

	This consistency guarantees us on the theoretical level that it 
is admissible to make this assumption, axiom 6, in our theory. The 
effectiveness of axiom 6 is solely checked through 
experiments and observations of actual physical phenomena. Some 
examples supporting this axiom on the experimental and observational 
level are given in Kitada 1994a, 1994b. We give below more comprehensive
 explanation of our idea about the effects of observations.

\MP

	To explain our idea more clearly, let us consider an observation of
 a local system consisting of two electrically neutral particles 
({\it e.g.}, neutrons) 1 and 2 with mass  $m_1,m_2$, 
interacting merely through gravitational force. In our theory,
 gravitation is generated by observation process through relativistic
 transformation of coordinates, and is not assumed {\it a priori}, 
{\it e.g.}, to be a Newtonian gravitational potential as a term in the
Hamiltonian of the local system under consideration. 
How can we then explain the gravitation between the two particles 1 and 2?
 We explain the gravitation as follows: Let the Hamiltonian of 
the local system of the two particles be
$$
H_0=h_{01}+h_{02},
$$
with $h_{01}$  and $h_{02}$  being the free Hamiltonians
 for the particles 
1 and 2 respectively. In this expression, there appears no gravitation.
 We decompose this system ``ideally"
 into two subsystems $L_1$  and $L_2$
  consisting of the particles 1 and 2 respectively. 

In the next paragraph,
 we use the terminology ``observation" as if
it were done actually about the positions and velocities of
the local systems $L_1$ and $L_2$. However, as remarked just
before axiom 6, these quantities considered below are intermediate
ones and are {\it not
observable directly}. The following description of
observations of such quantities should be interpreted as a 
description of a certain
``ideal" procedure of observation for getting a quantum mechanical
expression for gravitation. Therefore as a coordinate system
 of the observer's local system, we adopt the observer's
 coordinate system.

We suppose that we make an ``ideal" {\it observation} on particles 1 and 2
intermediately before the final observation of the scattering amplitude.
This ideal observation means that we make a relativistic transformation of
coordinates from the systems  $L_1$, $L_2$ of particle 1 and 2 to 
the observer's coordinates. Thus a gravitational field appears 
in this ideal observation along with this relativistic transformation
of coordinates. The classical general theory of relativity,
 namely axioms 4 and 5 together with an additional
 assumption as Einstein field equation, then yields that, in
the first order approximation, the Newtonian gravitational potential 
$Gm_1m_2/r$ is working between the two particles or the two subsystems  
$L_1$,  $L_2$, with distance $r$ moving with relative velocity $v$  at 
the time of this ``ideal" {\it observation}.
We assume that one of the particles, 
say particle 1, is stable with respect to the observer. We transform 
this classical observation into quantum mechanical 
statement, {\it i.e.} into a relativistically modified expression of
 the Hamiltonian $H$ above. The 
quantum mechanical evolution of the local system $L$ is given by
\beq
\nonumber
\exp[-itH]f&=&\exp[-ith_{01}]\exp[-ith_{02}]f\\
\nonumber
&\approx&\exp[-ith_{01}]f_1\exp[-ith_{02}]f_2
\ene
in the proper coordinate system of $L$,
 if we assume that $f\approx f_1\otimes f_2$  with $f,f_1,f_2$
  being the initial states of the local systems $L$, $L_1$, $L_2$
at the initial time $t=0$. In this ideal observation, the local systems 
$L_1,L_2$  are observed as possessing 
different times  $t_1,t_2$, respectively, in the observer's
 coordinate system of the observer. These times $t_1,t_2$
  are observed in accord with the relativistic 
transformation of coordinates in relation with the observer's space-time
 coordinates. If we denote the observer's local time by  
$t_O$, these times $t_1,t_2$  are expressed in terms of the observer's 
time $t_O$  as follows according to the special theory of relativity:
$$
t_1=t_O,\quad t_2=t_O\sqrt{1-(v/c)^2},
$$
where $c$ is the speed of light in vacuum. Thus, the evolution
$\exp[-itH]f$ is transformed, in the observer's coordinates,
as follows:
\beq
\nonumber
\exp[-itH]f&\mapsto&\exp[-it_Oh_{01}]f_1\exp[-it_O\sqrt{1-(v/c)^2}h_{02}]f_2\\
\nonumber
&\approx&\exp[-it_O\{h_{01}+\sqrt{1-(v/c)^2}h_{02}\}]f.
\ene
Here the subsystem Hamiltonians $h_{0j}$  are approximated by the
 relativistic classical energy
$$
h_{0j}\approx m_jc^2+p_j^2/(2m_j)\approx m_jc^2.
$$
The above evolution is, therefore, approximated by
$$
\exp[-itH]f\approx\exp[-it_O\{H_0-Gm_1m_2/r\}]f
$$
in actual observation, where we have used the classical energy conservation
 law to replace $p^2_2/(2m_2)=m_2v^2/2$  by  $Gm_1m_2/r$.
 This formula indicates that we can adopt the relativistic 
quantum mechanical Hamiltonian
\beq
H_0-Gm_1m_2/r
\ene
as an approximate relativistic Hamiltonian that is actually observed in the
 observation of the local system $L$ when looking into its sublocal systems
$L_1,L_2$. This is a typical example of our 
explanation strategy of the quantum mechanical relativistic phenomena
related with gravitation.
 We remark that this 
procedure consists of two processes: one is the {\it decomposition} 
of the observed system $L$ into 
some sublocal systems that are inquired by the observation, and another
 is the {\it relativistic 
transformation} of these subsystems' coordinates into the observer's
 coordinates{\footnote{The author is indebted to Dr. Masashi Oogami for the
idea to make the formulation of our treatment of gravity clear
in this way by introducing these two steps. The author is grateful to 
him for his kind permission for me to refer to 
this formulation in this paper.}}.

A concrete application of this consideration is the example
of the scattering of one neutron by two mirrors, which was
considered in Kitada 1994b, section 5. In this example, the gravitational
potential $-Gm_1m_2/r$ above is replaced by a uniform gravitation
$gL/c^2$. 
In accord with the remark
stated just before axiom 6 above,
the actually observable quantity in this example is
only the phase difference caused by the difference of the paths 
through which the two parts of the neutron pass.
The distance $r$ and
 the velocity $v$ used in the above are nothing but 
{\it theoretical apparatus}, which are {\it not observable directly}.

If the phenomena are concerned with electrical or other forces 
than gravity, the treatment is different as shown in the example 
of section 9 of Kitada 1994a. In this example, no gravity is assumed 
and the pure electrical forces are treated. In the case, the 
decomposition of the observed system is not possible, 
because the total Hamiltonian $H$ cannot be decomposed as a sum 
of mutually commuting $h_{0j}$ as in the above. Instead, one has 
to treat the total Hamiltonian $H$ to deduce the differential
 cross section (14) of Kitada 1994a from a purely 
quantum mechanical consideration. Only after this procedure, the
relativistic effect of observation should be taken into the
 expression (14) to reach (17) and (19) in p.297 of Kitada 1994a.

If the observed system includes both of gravity and other forces,
 the order of treatment is as follows: First, consider the 
quantum mechanical Hamiltonian to deduce some formulae
 which explain the
 purely quantum mechanical phenomena about the system. Then,
 considering the relativistic effect of the observation, transform
 the coordinates of the system to the coordinates of the observer's
 system. Then gravitation is automatically included in the final
 expression of the formulae.

These examples describe our idea of treating the relativistic
phenomena like gravitation as the effect or the consequence of 
the observation.
This will be discussed in more details elsewhere.
\BP

The above axiom 6 asserts, also, that the {\it actually observed values}
 of a quantum mechanical quantity must be {\it classical}. In other words,
 the quantum mechanical quantity is measured, in actual observation, 
as a classical quantity. Thus this axiom together with our other 
postulates resolves the fundamental problem of the consistency of the
 Copenhagen interpretation that the quantum mechanics must be tested
 through the experiments that are described by the use of classical
 physics (see Jammer 1974, section 4.2). This is another 
implication of our formulation. We also remark that the paradox 
presented by Einstein-Podolsky-Rosen 1935 can be resolved by our
 formulation (see Kitada 1994a, section 8), where the usual inconsistency
 between non-locality of quantum mechanics and local causality of 
relativistic theory is treated without contradiction, as mutually
 independent aspects of nature.

	This axiom 6, by its nature, gives a certain relation between
 quantum mechanical and classical views so that the so-called relativistic
 quantum mechanical phenomena could be explained as in 
the above explanation. We repeat that,
 as far as we know, there is no theory that can explain relativistic
 quantum mechanical phenomena in any consistent way: The 
present quantum field theories, which look like successful at least in
 explaining some examples like Lamb shift, have not been constructed 
consistently or have never been proved to be 
consistent. See, {\it e.g.}, Streater's paper in Brown-Harr\'e 1990, 
and Fr\"ohlich 1982. See also Dyson 
1953, Jaffe 1965, and Calan-Rivasseau 1982 for the proof that the series
 giving Lamb shift diverges. These results indicate a theoretical
 contradiction of quantum field theories which give 
predictions for the Lamb shift in `outstanding' accuracy
 (see Kinoshita {\it et al}. 1983) by ``theoretical" calculations based on
 the approximation up to the 6-th or 8-th order of the 
``divergent" series.

	We conclude this section by remarking that if the state
 of a local system $L$ is a bound state of a local Hamiltonian $H_L$ 
 corresponding to a local system $L$, then that state is not observable,
 insofar as the observer observes that local system. For if
 the state function $\psi_L$ of the system $L$ is a bound state of $H_L$,
 it cannot emit any light or information outside, hence cannot be 
 observable by any observer. However, if the observer changes its view,
 {\it e.g.}, if it observes a larger system $L'$ including the system $L$,
 then there is a possibility that the observer sees the system $L$,
 because the larger system $L'$ may be in a scattering state
of the larger Hamiltonian $H_{L'}$,
 and as one of its subsystems, $L$ may be also observable.
 In general, what we see are dependent on what we intend to see,
 or in other words it is dependent on which part of 
the universe we are trying to see.
 We will try to clarify this point some more in the next section.

\vskip24pt

\large
\noindent
{\bf VII. Observation, which solves the current measurement problem}
\normalsize
\vskip12pt

\noindent
Let us begin with a seemingly contradictory example. When one is 
observing a stable hydrogen atom, why is it stable even if one is 
observing it? If the observer is inquiring into the hydrogen atom 
dividing it into an electron and a nucleus, they should obey classical 
physics by our axiom 4, hence the system has to decay according to 
the classical electromagnetic dynamics, with the electron going down 
into the nucleus, emitting photons. Would this not damage the 
quantum mechanical explanation of stability of atoms, getting the 
situation back to that of 19th century? 

        Our answer to this paradox is that, when one sees that the 
hydrogen atom is stable, the observer is observing the hydrogen atom 
as a local system consisting of the electron and the nucleus, but does 
not make any inquiry into its inside. In fact the stability of an atom
 is explained in QM as a property of the local Hamiltonian for the 
local system of the atom, precisely speaking as the semiboundedness 
property of that Hamiltonian, which is violated in classical mechanics.
 On the contrary, when one observes into the inside of the atom, 
the observer stimulates the atom by some method, {\it e.g.} by hitting 
the atom by other particles, to get some information on the 
internal structure of the atom, hence this observation would destroy
 the atom, if the observation would be successfully done to get the
 required information. In this case the objects of the observation, 
{\it e.g.} the electron, the nucleus, and the other particles used to 
make the observation, follow classical mechanics during the 
observation to result in the destruction of the atom along the
 scenario given by classical electromagnetic theory.
 This would give a first explanation of the decaying process of
 the atom in a classical mechanical way, which has been a mystery
 in the context of quantum mechanics. 

       We remark that the explanation given here is the same one as
  the ones given in sections 7 and 8 of Kitada 1994a, where
we have treated, {\it e.g.}, EPR paradox.

       As another but more fundamental problem, which has been the crux 
of the current measurement problem, let us see the problem of the
 so-called ``collapse" or ``reduction" of wave functions, which is
 usually supposed to occur at the time of measurement. In the
 ordinary formulation of quantum mechanics, there is a gap between
 the quantum mechanical description as a wave function of the state
 under consideration and the actually observed values about the state.
 The former evolves according to an equation of motion, {\it i.e.}
 Shr\"odinger equation, but the actually observed values are considered
 as a result of ``collapse" of the wave function to an eigenstate
 of the Hamiltonian which describes the system under consideration.
 We borrow an explanation of this problem from Auyang, 1995,
 section 4 The Quantum Measurement Problem (page 22), which describes
 the exact nature of the problem:

\MP

\leftskip24pt
\rightskip24pt

\noindent
Quantum mechanics gives two descriptions that differ in nature, subject
 matter, and treatment. The characteristics described by state vectors
 are nonclassical; they are irreducibly complex and strangely entangled
 when expressed in classical terms. Besides its nonclassicality,
 the state-vector description is decorous; it is essentially of single
 systems evolving according to an equation of motion. The description
 offered by the statistics of eigenvalues is just the opposite. The
 characteristics it describes are classical and familiar. But besides
 its classicality, it has the appearance of bastard; it applies not to
 single systems but only to ensembles, and it is the result of 
``collapse" of nonclassical characteristics. The crux of the problem
 is that the quantum mechanics provides no substantive correlation
 between the two descriptions. The only relation between them is
 formal and abstract. It is provided by the observable, whose
 eigenstates contribute to the state description and whose eigenvalues
 to the statistical description.

    The stepchild treatment accorded to the classical description by a
 theory that many interpreters claim to be universal and fundamental
 is regrettable, for from a broad perspective that includes but is not
 limited to quantum mechanics, these characteristics are objective.
 Classical characteristics are realized in the physical instruments
 used in quantum experiments, and they are subjected to the laws of
 classical physics. In the usual understanding, the classical and
 quantum descriptions are said to be connected in measurements. However,
 we do not have even a marginally satisfactory account of the
 measurement process. This is known as the quantum measurement problem.

\leftskip0pt
\rightskip0pt

\MP

       As described in this quotation, the point of the problem is
 the lack of procedure which explains the collapse {\it inside the framework
 of quantum mechanics}. A rigorous theory of measurements developed by
 P. Busch, P. J. Lahti, and P. Mittelstaedt quoted in the same section of
 Auyang 1995 shows that the difficulty of the problem lies in
 {\it objectification}, which demands that the instrument realizes
 a definite eigenvalue at measurement. Many have tried to respond to this
 demand inside the framework of quantum mechanics. We repeat the words
of P. Busch et. al
 quoted in Auyang 1995: ``One would expect, and most researchers on the
 foundations of quantum mechanics have done so, that the problem of
 measurement should be solvable {\it within} quantum mechanics."  

       Our stand to this problem is to leave the word 
``observation" or ``measurement" as an undefined term of our theory,
 as we have stated in section VI, so that the term cannot be
 analyzed and explained inside the framework of our theory.
 We do not give any explanation to the ``collapse" and the measurement
 process within our theory, just similarly to Newton who left the
 term ``gravity" as a mere name, to whose cause he did not give explanation.
 Instead we regard observation as an actual deed we do in usual life, and
 give a procedure of calculation of the scattering amplitudes which
 are {\it only} the quantities observable in actual
 quantum mechanical experiments. 

       What is important to know is that there are things which cannot
 be known to human beings. The modern science tends to think that it
 can explain everything in the universe, as the attempts to ``Theory of
 Everything" indicate. However, we already know that there are things
 which cannot be known to human beings even in such a region of pure reason
 as mathematics, where K. G\"odel 1931 showed that there are infinitely many
 number of propositions in a mathematical theory that includes natural
 numbers, the proofs of which and whose negation cannot be obtained in
 the framework of that theory. Those who are careful enough to be aware of
 that such a deep insight to the restriction of human ability of getting
 things has already been found at such an early stage of the 20th century,
 would agree with our stand that we do not try to explain everything around
 us and we leave the observation as actual activities of human beings which
 cannot be explored and analyzed any more.

        Another point which should be stated about the current attempts
 to explain the measurement process is that these attempts have tried
 to explain that the instrument realizes a definite eigenvalue, but failed.
 This failure is not at all a problem from our standpoint.
 These attempts just
 explain the interaction between the instrument and the object
 in the framework of quantum mechanics. In our theory, this explanation is
 regarded as just an explanation of the quantum mechanical ``scattering
phenomenon" between the instrument and the object, and it is no wonder
 that it does not result in the ``collapse" of wave function to give
 a definite eigenvalue.

\MP

        These two examples show that there are two categories in the
 current measurement problems: One category arises from the misuse of the
 quantum mechanics and classical mechanics. In the stability problem of
 atoms, one is confused 
in the point where one should put a borderline between the
 quantum mechanical view and classical mechanical view. The problem is
 solved when one finds which local systems the observer is observing at each
 moment. Depending on the local systems that the observer observes, the view
 which the observer gets from nature is different. Even when one thinks
 that one is observing the same object, there are large possibility in
 which local systems are selected in each observation. According to
 this selection, the looks of things change from one to another.
 In the next section we will see another example of this kind.

       Another category of problems has more deep root, which
 concerns with the human ability and inability of knowing things.
 This kind of borderline about our ability has already been set at
 the end of section III of the present paper. It is the statement
 of the unobservability
 of quantum mechanical nature of the universe. Although quantum mechanical
 nature of the universe
is unobservable, we,  as one of local systems inside
 the universe, can perceive it,
 since the change outside us influences us and {\it vice versa},
 so that the total universe is stationary. This is just the same situation
 as in G\"odel's incompleteness theorem, where although there are
 undecidable propositions whose proofs cannot be obtained by hands,
 the correctness of those propositions can be known from what they mean
in the larger context.
 The universe is not an existence reachable by hands from human beings,
 but it tells itself to us by its eternal and total nature.

\vskip24pt

\large
\noindent
{\bf VIII. Hubble's Law}
\normalsize
\vskip12pt

\noindent
As another example of the applicability of our framework to relativistic
 quantum mechanical phenomena, we give in this section some outline of
 the explanation of Hubble's law, which was treated in Kitada 1994b.
 We will also touch on another possible explanation of the law.

	Hubble's law is a phenomenon which appears in observing
 the light emitted
 from astronomical existence such as stars, galaxies, etc. The emission of
 light is a quantum mechanical phenomenon which could be treated in our
 framework as in Kitada 1994a, section 11-(2). The observation on the earth
 of this emission of light from stars can be explained as a classical 
relativistic observation, by our assumption on observation stated in 
the previous section.

	The mathematical treatment was given in Kitada 1994b, section 6. 
Here we explain why our stationary static universe admits an 
`expansion' of this kind.

	The `expansion' in the large scale, as a result of our general
 principles of relativity and the additional assumption, {\it i.e.}
 the Einstein's field equation, is an {\it observational} fact.
 This is no 
contradiction with our assumption of stationary universe,
 because the stationary nature of the universe is a quantum mechanical
 one and {\it unobservable}. 

	More specifically, in explaining the `expansion,' one has to
 adopt a fixed observer's space coordinates, which are in this case the
 `comoving, synchronous space coordinates,' while, 
as the time coordinate, we assume that it `slices' the spacetime by a
 one parameter family of spacelike surfaces (see, {\it e.g.},
 Misner-Thorne-Wheeler 1973, Chap.27). The `expansion' appears 
only after these spacetime coordinates are fixed. This choice looks natural,
 but there are infinitely many other possibilities of the choice of
 coordinates. The fact that the Hubble's law can 
be explained by the general relativity means no more than that
 our usual choice of coordinates accidentally coincides with this
 `natural' choice of coordinate system of comoving, synchronous 
one. Or rather, reversely saying, the synchronous coordinates are 
chosen so that they coincide with our daily choice of coordinates
 in our astronomical observation.

	This is no contradiction with our definition of stationary universe
 as a quantum mechanical eigenstate of a total Hamiltonian $H$. This
 Hamiltonian and therefore the total universe $\phi$  are hidden
 behind the observable phenomena. They appear only through some 
appropriately organized classical observations. Not all the 
quantum mechanical world can be observed, but only a few of them appear
 before our eyes through some well-designed 
experiments or measurements. One of them is Hubble's law, which is 
a classical observation of the quantum mechanical emission of light
 from stars, though it mainly reveals the classical aspect 
of the universe, except for the spectrum structure of the light emitted
 from stars.

	Observation and the true world can be different; which is not 
contradictory at all. The `big bang' as claimed as a logical consequence
 of the Hubble's law or expansion might be 
natural in its classical mechanical logic. 
However, the beginning of the universe is no more than 
an imagination, which cannot be seen actually, even if one had a 
`time machine,' for the machine itself would contract at an initial
 point of the big bang when it goes back to the time of big bang, 
and it could not report the big bang phenomenon to the age when the
 machine started. Big bang cannot be an object of science in this sense
 of nonreproducibility of its phenomenon. 

	More precisely speaking, our assertion is as follows: What looks 
like the expansion depends on the choice of coordinates. A mathematical
 treatment in Kitada 1994b shows that the expansion is a one in a 
`virtual' 4-dimensional Euclidean space, that is different from the one in
the curved Riemannian manifold, in which we live according to the general 
relativity. This Euclidean space is borrowed from the outside of the
 Riemannian manifold in order to visualize the `expansion' 
phenomenon. The mathematical fact is that the Robertson-Walker metric
 can be rewritten in a certain form, which can be `interpreted'
 as this visualization of expansion. The actual result is a Riemannian
 geometry, whose interpretation as `expansion' is only a convenient 
one.

	Here we note that there is another possibility to explain the Hubble's
 law, which fits the actual cosmological observations in more precise
 manner. As we have mentioned at the end of section V, one can assume
 another field equation than the Einstein's, the Hoyle-Narlikar field 
equation which has a conformally equivalent solution to the Einstein field
 equation, whereas the interpretation given by Arp 1993 requires the
 creation of matters continuously at all points in the universe. This looks,
 at a first glance, contradicting our assumptions of quantum mechanics, but,
as has been stated in section V, the Hoyle-Narlikar filed equation is a
 {\it classical} equation consistent with the two principles of general 
relativity which are concerned with the {\it observed}
facts, hence, that field equation can be adopted 
as our basic field equation to determine the 
metric $g_{\mu\nu}$  without inconsistency with the constancy of the 
{\it quantum mechanical}
 masses of particles. The adoption
 of this type of field equation then gives a more 
preferable explanation than the usual `expansion' explanation, of the
 astronomical observations, {\it e.g.},
 of quasars redshift, of the nearby stars redshift in and around the 
Milky Way, and so on (see Arp 1993).

	In this way, our theory gives a flexible {\it framework}
 for the explanation
 of cosmological phenomena. The reader may, however, ask: If big bang 
is not an object of science, how can the stationary total universe in the
 present theory be justified without having the observability? The 
answer may be obvious from what we have mentioned, but we repeat it as a 
summary: if one assumes this universe,
 then one can explain some phenomena 
from cosmological size as Hubble's law to the human size as the experiment
 of the interference of a neutron (see Kitada 1994b, section 5) with much
 flexibility. Also some microscopic phenomena can be explained 
(see Kitada 1994a, sections 7, 9). We emphasize that these are 
explained in {\it one framework}
 in a {\it consistent way}, differently from the
 existing explanations. We have also suggested a possibility 
of explanation of Lamb shift in Kitada 1994a, section 11-(2),
 in our framework, and a possibility 
of explaining the stability of galaxies, etc., as a quantum mechanical
 property of (approximate) eigenstates of local Hamiltonians associated 
with local systems, without appealing to the `dark matter' which has not
 been observed or is considered as invisible matter. Our answer to the
question above, therefore, is the unified manner of our framework that
 can explain some actual relativistic quantum mechanical phenomena 
as well as some fundamental cosmological problems, {\it e.g.},
 the recently noticed ``cosmological conflict" between the 
`ages' of the universe and the stars 
(see M. J. Pierce et al. 1994, W. L. Freedman et al. 1994, 
C. J. Hogan 1994, G. H. Jacoby 1994, B. Chaboyer et al. 1996).

\vskip24pt

\large
\noindent
{\bf IX. Discussions}
\normalsize
\vskip12pt

\noindent
{\bf 1.}\quad
  Our leading principle of the construction of the theory has been
 the logical consistency of the theory. QM and GR has been introduced as
 mutually independent aspects of nature, by orthogonalizing these two
 theories. Then we have gotten a complete freedom to make an 
arbitrary assumption about the relation between these theories, as far as
 the assumption is consistent with QM and GR. We have chosen one possible
 assumption, axiom 6, so that it would explain the actual relativistic
 quantum mechanical phenomena. The validity of this axiom 6 can 
be checked by experiments and observations of actual physical phenomena.
 This requirement can be called again a requirement for {\it consistency}
 with the actual physical phenomena.
\MP

\noindent
{\bf 2.}\quad
  Our notion of time formulated in the above is an alternative for,
 or an intermediate notion between Newton's and Einstein's notions of time,
 in the following three points: 1) we define the local time as a measure
 of motion, as a substitute to Newton's absolute time defined as a clock 
of the motion of bodies; 2) we can use the local time as the Einstein's 
(general) relativistic proper time at the center of mass of a local
 system, so that we can recover the classical relativistic world; and 3)
 we can maintain the notion of absolute time dominating the total 
universe, although our notion of the absolute 
or total time works in quite a contrary manner to Newton's absolute 
time. 

\vskip38pt

\large

\noindent
{\bf Acknowledgements}

\normalsize

\BP

\noindent
The present work was started by the request
 from Mr. C. Roy Keys of Apeiron to write a paper which would be readable for
 non-specialists to understand Kitada 1994a.
 He helped us in various ways
 in the course of the preparation of the present version after H. K. wrote the
 first manuscript, which was given negative responses from
 some referees.  He, nevertheless,
 finds certain meanings in the work, and has encouraged us ungrudgingly.

When H. K. wrote the first manuscript of ``Theory of Local Times,"
 Kitada 1994a, which is the starting one of the present series of works,
Dr. Izumi Ojima played an important role in the publication of the paper,
without which the present work is not. He read thoroughly the first
 manuscript of ``Theory of Local Times," and gave H. K. a comprehensive list
 that contains his questions and criticisms. Owing to this list, H. K. could
 avoid unnecessary descriptions in the first manuscript of 
``Theory of Local Times," which might have
 led the reader to the out of context of the paper.

H. K. is also indebted to Dr. Masashi Oogami, who is a physicist and 
communicated with H. K. on line, discussing mainly the treatment of the
relativistic phenomena including gravitational ones in the present framework.
 H. K. owes many points in section VI to the discussions with him.

       H. K. had an opportunity at almost the final stage of the 
work to have Mr. Jacob Schach M{\o}ller at Dept. Math. Sci., Univ. 
Tokyo as a visitor from Aarhus Univ., Denmark, and to have had 
chances to discuss on the work. Section VII has been added 
owing to the discussion with him. H. K. expresses appreciation 
to him for the discussion.

We also thank Mr. Peter Unwin, who 
gave us a chance to discuss on CompuServe, and also encouraged H. K. 
at the early stage of the work.

\vskip38pt

\large
\noindent
{\bf References}

\BP

\small

\F
H. Arp, 1993, ``Fitting theory to observation -- from stars to cosmology,"
 in Progress in New Cosmology, Plenum Press, New York-London.

\F
A. Ashtekar, J. Stachel (eds.), 1991, Conceptual Problems of Quantum Gravity,
 Birkh\"auser, Boston-Basel-Berlin.

\F
S. Y. Auyang, 1995, How is Quantum Field Theory Possible? Oxford
Univ. Press, New York-Oxford.

\F
D. Bohm, B. J. Hiley, 1993, The Undivided Universe, An ontological 
interpretation of quantum theory, Routledge, London and New York.

\F
H. R. Brown, R. Harr\'e (eds.), 1990, Philosophical Foundations of Quantum
 Field Theory, Clarendon Press, Oxford.

\F
C. Calan, J. Rivasseau, 1982, ``The perturbation series for   
field theory is divergent," Commun. Math. Phys. {\bf 83}, 77-82.

\F
B. Chaboyer et al., 1996, ``A lower limit on the age of the universe,"
 Science, 271, 957-961.

\F
R. Descaretes, B. Spinoza, G. W. Leibniz, 1960, The Rationalists, 
Descartes, Spinoza, Leibniz, 
Doubleday, New York-London-Toronto-Sydney-Auckland.

\F
P. A. M. Dirac, 1958, The Principles of Quantum Mechanics, 4-th ed.,
 Clarendon Press, Oxford.

\F
F. J. Dyson, 1953, ``Divergence of perturbation theory in quantum
 electrodynamics," Phys. Rev. {\bf 75}, 486.

\F
A. Einstein, 1905, ``Zur Elektrodynamik bewegter K\"orper," 
Annalen der Physik, {\bf 17}, 891-921.

\F
A. Einstein, 1916, ``Die Grundlage der algemeinen Relativit\"atstheorie,"
 Annalen der Physik, Ser.4, {\bf 49}, 769-822.

\F
A. Einstein, 1920, Relativity, The Special \& The General Theory,
 Tr. R. W. Lawson, Methuen \& Co. LTD.

\F
A. Einstein, B. Podolsky, N. Rosen, 1935, ``Can quantum mechanical
 description of physical reality be considered complete?" Phys. 
Rev. {\bf 47}, 777-780.

\F
V. Enss, 1986,  ``Introduction  to  asymptotic observables for
 multiparticle quantum scattering," in Schr\"odinger Operators, 
Aarhus 1985, ed. E. Balslev, Lect. Note in Math. {\bf 1218}, Springer-Verlag,
pp.61-92.

\F
R. P. Feynman, 1948, ``Space-time approach to non-relativistic quantum
 mechanics," Rev. Modern Phys., {\bf 20}, 367-387.

\F
W. L. Freedman et al., 1994, "Distance to the Virgo cluster galaxy 
M100 from Hubble Space Telescope observations of Cepheids," Nature, 
{\bf 371}, 757-762.

\F
J. Fr\"ohlich, 1982, ``On the triviality of $\lambda\Phi_d^4$
  theories and the approach to the critical point in $d\le 4$
  dimensions," Nucl. Phys. {\bf B 200 [FS4]}, 281-296.

\F
K. G\"odel, 1931, ``\" Uber formal unentsceidebare S\" atze der Principia
 mathematica und verwandter Systeme I," Monatshefte f\" ur Mathematik
 und Physik, (or Kurt G\" odel Collected Works, Volume I, Publications
1929-1936, Oxford University Press, New York, Clarendon Press, Oxford,
 1986, pp.144-195).

\F
J. B. Hartle, 1993, ``The spacetime approach to quantum mechanics," 
in Proceedings of the International Symposium on Quantum Physics and 
the Universe, Waseda University, Tokyo, Japan, August 23-27, 
1992. (gr-qc/9210004)

\F
W. Heisenberg, 1925, ``\"Uber quantentheoretische Umdeutung kinematischer
 und mechanischer Beziehungen," Zeitschrift f\"ur Physik, {\bf 33}, 879-893.

\F
C. J. Hogan, 1994, ``Cosmological conflict," Nature, {\bf 371}, 374-375.

\F
C. J. Isham, 1993,  ``Canonical quantum gravity and the problem of time,"
 in Proceedings of the NATO Advanced Study Institute, Salamanca, 
June 1992, Kluwer Academic Publishers.  (gr-qc/9210011)

\F
G. H. Jacoby, 1994, ``The Universe in crisis," Nature, {\bf 371}, 741-742.

\F
A. Jaffe, 1965, ``Divergence of perturbation theory for Bosons," Commun.
 Math. Phys. {\bf 1}, 127-149.

\F
M. Jammer, 1974, The Philosophy of Quantum Mechanics, The interpretations
 of Quantum Mechanics in Historical Perspective, John Wiley \& Sons, Inc.,
 New York.

\F
T. Kinoshita, W. B. Lindquist, 1983, ``Eighth-order magnetic moment 
of the electron," Phys. Rev. {\bf D27}, 867.

\F
H. Kitada, 1980, ``On a construction of the fundamental solution
for Schr\"odinger equations," J. Fac. Sci. Univ. Tokyo, Sec. IA,
{\bf 27}, 193-226.

\F
H. Kitada, 1994a,  ``Theory of local times," Il Nuovo Cimento, 
{\bf 109 B, N. 3}, 281-302. (astro-ph/9309051)

\F
H. Kitada, 1994b, ``Theory of local times II. Another formulation and 
examples," preprint. (gr-qc/9403007)

\F
C. W. Misner, K. S. Thorne, J. A. Wheeler, 1973, Gravitation, W. H. Freeman
 and Company, New York.

\F
J. V. Narlikar, 1977, ``Two astrophysical applications of conformal 
gravity," Ann. of Phys. {\bf 107}, 325-336.

\F
I. Newton, 1962, Sir Isaac Newton Principia, Vol. I The Motion of Bodies,
 Motte's translation Revised by Cajori, Tr. Andrew Motte ed. 
Florian Cajori, Univ. of California Press, Berkeley, Los Angeles, London.

\F
M. J. Pierce et al., 1994, ``The Hubble constant and Virgo cluster 
distance from observations of Cepheid variables," Nature, {\bf 371}, 385-389.

\F
M. Sachs, 1986, Quantum Mechanics from General Relativity, Reidel.

\F
M. Sachs, 1988, Einstein versus Bohr, Open Court.

\F
E. Schr\"odinger, 1926, ``Quantisierung als Eigenwertproblem 1, 2,"
 Annalen der Physik, {\bf 79}, 361-376, 489-527.

\F
W. G. Unruh, 1993, ``Time, gravity, and quantum mechanics," preprint. 
(gr-qc/9312027)

\end{document}